\documentclass[prb,twocolumn,superscriptaddress]{revtex4}
\usepackage{amssymb}
\usepackage{graphicx,amsmath}
\usepackage{bm}
\usepackage{times}

\begin{document}

\title{Inverse Landau-Zener-St\"{u}ckelberg problem for qubit-resonator
systems}
\author{S. N. Shevchenko}
\email{sshevchenko@ilt.kharkov.ua}
\affiliation{B.Verkin Institute for Low Temperature Physics and Engineering, Kharkov,
Ukraine}
\affiliation{RIKEN Advanced Science Institute, Wako-shi, Saitama, Japan}
\author{S. Ashhab}
\affiliation{RIKEN Advanced Science Institute, Wako-shi, Saitama, Japan}
\affiliation{Department of Physics, The University of Michigan, Ann Arbor, Michigan, USA}
\author{Franco Nori}
\affiliation{RIKEN Advanced Science Institute, Wako-shi, Saitama, Japan}
\affiliation{Department of Physics, The University of Michigan, Ann Arbor, Michigan, USA}

\begin{abstract}
We consider theoretically a superconducting qubit - nanomechanical resonator
(NR) system, which was realized by LaHaye et al. [Nature \textbf{459}, 960
(2009)]. First, we study the problem where the state of the strongly driven
qubit is probed through the frequency shift of the low-frequency NR. In the
case where the coupling is capacitive, the measured quantity can be related
to the so-called quantum capacitance. Our theoretical results agree with the
experimentally observed result that, under resonant driving, the frequency
shift repeatedly changes sign. We then formulate and solve the inverse
Landau-Zener-St\"{u}ckelberg problem, where we assume the driven qubit's
state to be known (i.e. measured by some other device) and aim to find the
parameters of the qubit's Hamiltonian. In particular, for our system the
qubit's bias is defined by the NR's displacement. This may provide a tool
for monitoring of the NR's position.
\end{abstract}

\date{\today}
\maketitle

\section{Introduction}

Nanoelectromechanical systems have recently attracted attention because of
both possible applications (e.g. in sensing) and interest in fundamental
quantum phenomena in mesoscopic systems.\cite{Blencowe05} Particularly
interesting is the coupling of the mechanical motion of a nanomechanical
resonator (NR) to an electric mesoscopic system. A few examples are carbon
nanotube NRs coupled to electron transport \cite{Lassagne09} and a metallic
NR coupled to an $LC$ tank circuit \cite{Sillanpaa09}. It was proposed
theoretically that for sensing and controlling the NRs, superconducting
few-level circuits (qubits)\cite{YouNori} can be effectively used. \cite%
{Irish03, Martin04} For example this approach was applied in the
demonstration of the ground state of a high-frequency piezoelectric
dilatational resonator coupled to a superconducting phase qubit.\cite%
{OConnell10}

Successful coupling of a NR (a suspended silicon nitride beam) to a charge
qubit allowed LaHaye \textit{et al.}~[\onlinecite{LaHaye09}] to demonstrate
both ground-state measurement and excited-state spectroscopy as well as
Landau-Zener-St\"{u}ckelberg (LZS) interferometry of the qubit. The
spectroscopy was performed with weak driving, where the position of the
resonance gave the information about the qubit levels. In the regime of
strong driving, where the qubit's evolution experiences repeated LZS
transitions at the avoided crossing, the resulting interference is
visualized in the LZS interferograms [\onlinecite{SAN}]. The LZS
interferometry was demonstrated on superconducting qubits probed by
different methods (see Ref.~[\onlinecite{SAN}] and references therein), as
well as studied for other different physical realizations of strongly-driven
two-level systems in Refs.~[\onlinecite{MISC}].

In the work by LaHaye \textit{et al.}, Ref.~[\onlinecite{LaHaye09}], the
NR's frequency shift was used for monitoring the qubit's state. For the
theoretical description of the NR-qubit system, the perturbation-theory
procedure developed in Ref.~[\onlinecite{Irish03}] was used. The theory says
that the NR's frequency shift $\Delta \omega _{\mathrm{NR}}$\ is negative
for the qubit in the ground state and zero when the two qubit states are on
average equally populated under the periodic driving. This allowed to
describe the ground-state and low-amplitude spectroscopy measurements.\cite%
{LaHaye09} However, this theory does not explain the experimentally observed
sign changes of $\Delta \omega _{\mathrm{NR}}$\ in the strong-driving
regime, where the frequency shift becomes positive.

In this work we consider the NR-qubit system semi-classically. Within this
approach, we describe the qubit as a quantum system coupled to a classical
resonator, with the oscillation-energy quantum much smaller than the thermal
energy, $\hbar \omega _{\mathrm{NR}}\ll k_{B}T$. Note that such a
semi-classical approach was successful for the description of most phenomena
related to atom-light interaction.\cite{Delone85}

The impact of the qubit on the resonator's frequency shift can be described
in terms of the so-called quantum capacitance, as studied for the qubits in
Refs.~[\onlinecite{Sillanpaa05, Duty05}]. The quantum capacitance is defined
as the derivative of the average charge on the qubit with respect to the
applied voltage. The charge can then be related to the charge-qubit
occupation, the derivative of which (under resonant driving) exhibits sign
changes. Similar sign-changing response under strong driving was recently
studied for qubits probed by an $LC$ (tank) circuit for capacitive coupling
\cite{Sillanpaa06, Paila09} as well as for inductive coupling \cite%
{Shnyrkov06, Shevchenko08}. Thus, in the first part of this work
(Section~II) we study the situation where the strong-driving qubit's state
is probed by the NR.

\begin{figure}[t]
\includegraphics[width=7 cm]{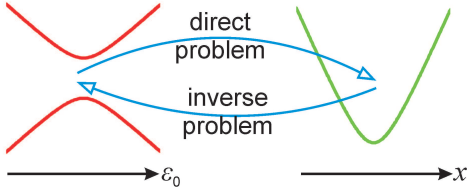}
\caption{(Color online) Schematic representation of the formulated problems
for direct and inverse interferometry. The red curves on the left represent
the bias-dependent energy levels of the qubit, and the green parabola on the
right shows the potential energy of the (classical) resonator. In the direct
problem, the resonator is used to probe the state of the qubit. In the
inverse problem, the response of the qubit to external driving is used to
infer the state of the resonator.}
\label{Scheme0}
\end{figure}

In Section~III, we formulate the inverse problem. There, we are interested
in the influence of the NR's state (its position) on the qubit's state. We
graphically demonstrate the formulation of the problem for the direct and
inverse interferometry in Fig.~\ref{Scheme0}. There, the two-level system
represents a qubit with control parameter $\varepsilon _{0}$; the parabola
represents the resonator's potential energy as a function of the
displacement $x$. Thus, in the first part of our work (Sec.~II) we deal with
the direct problem, where the influence of the qubit's state on the
resonator is studied.

The second part of this work (Secs.~III\ and IV) is devoted to the inverse
problem, where we study the influence of the resonator's state on the
qubit's state. Measuring the latter is an alternative method for defining
the NR's displacement. This approach can be related also to other inverse
problems for two-level systems, as studied in Refs.~[%
\onlinecite{Garanin02,
Berry09, Burgarth09}]. Generalization of the results can also be applied to
other quantum systems for which the problem of defining the Hamiltonian's
parameters with given system's state was studied in Ref.~[\onlinecite{TerNak}%
]. In Section IV we demonstrate how the inverse problem can be solved for
different driving regimes in a generic two-level system, and we comment on
the possibility of applying this technique for superconducting qubit-NR
systems.

\section{Charge qubit probed through the quantum capacitance}

\begin{figure}[t]
\includegraphics[width=8 cm]{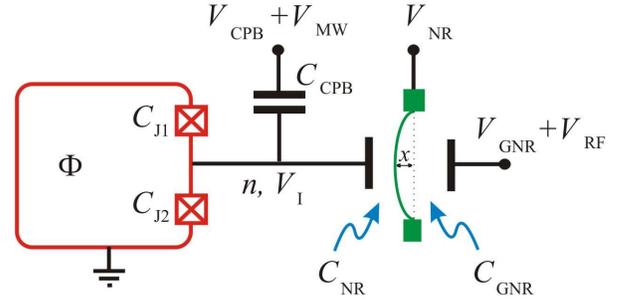}
\caption{(Color online) Schematic diagram of a split-junction charge qubit
coupled to a nanomechanical resonator. The charge qubit (shown in red) is
biased by the magnetic flux $\Phi $ and the dc+$\protect\mu $w voltage, $V_{%
\mathrm{CPB}}+V_{\mathrm{MW}}$, to which it is coupled through the
capacitance $C_{\mathrm{CPB}}$. The qubit is coupled to the NR (shown in
green) through the capacitance $C_{\mathrm{NR}}$. The NR is biased by a
large dc voltage $V_{\mathrm{NR}}$; its state is controlled and measured by
applying the dc and rf voltages between the gate and the NR, $V_{\mathrm{GNR}%
}$ and $V_{\mathrm{RF}}$, through the capacitance $C_{\mathrm{GNR}}$. The
NR's motion is described by the displacement at the midpoint $x$.
Capacitances form the island (Cooper-pair box) with the total capacitance $%
C_{\Sigma }$, voltage $V_{I}$ and charge $-2en$.}
\label{Fig:scheme}
\end{figure}

The split-junction charge qubit (also called Cooper-pair box and shown in
red in Fig.~\ref{Fig:scheme}) consists of a small island between two
Josephson junctions. The state of the qubit is controlled by the magnetic
flux $\Phi $ and the gate voltage $V_{\mathrm{CPB}}+V_{\mathrm{MW}}$. Here $%
V_{\mathrm{CPB}}$ is the dc voltage used to tune the energy levels of the
qubit and $V_{\mathrm{MW}}=V_{\mu }\sin \omega t$ is the microwave signal
used to drive and manipulate the energy-level occupations. The Cooper-pair
box is described in the two-level approximation by the Hamiltonian in the
charge representation (see e.g.~Ref.~[\onlinecite{LaHaye09}] and Appendix~A)%
\begin{equation}
H(t)=-\frac{\Delta }{2}\sigma _{x}-\frac{\varepsilon _{0}}{2}\sigma _{z}-%
\frac{A\sin \omega t}{2}\sigma _{z}.  \label{H(t)}
\end{equation}%
Here the tunnel splitting $\Delta $ is equal to the Josephson energy $E_{%
\mathrm{J}}$, which is controlled by the magnetic flux $\Phi $%
\begin{equation}
\Delta \equiv E_{\mathrm{J}}=E_{\mathrm{J}0}\left\vert \cos (\pi \Phi /\Phi
_{0})\right\vert .  \label{Delta}
\end{equation}%
The charging energy and the driving amplitude are given by%
\begin{equation}
\varepsilon _{0}=8E_{\mathrm{C}}(n_{\mathrm{g}}-1/2)\text{, \hspace{0.5cm} }%
A=8E_{\mathrm{C}}n_{\mathrm{\mu }}\text{,}  \label{e0}
\end{equation}%
where the Coulomb energy $E_{\mathrm{C}}=e^{2}/2C_{\Sigma }$ is defined by
the total island's capacitance $C_{\Sigma }=2C_{\mathrm{J}}+C_{\mathrm{CPB}%
}+C_{\mathrm{NR}}$, defined with the notation $2C_{\mathrm{J}}\equiv C_{%
\mathrm{J}1}+C_{\mathrm{J}2}$; the dimensionless driving amplitude is $%
n_{\mu }=C_{\mathrm{CPB}}V_{\mu }/2e$; the dimensionless polarization charge
$n_{\mathrm{g}}=n_{\mathrm{NR}}+n_{\mathrm{CPB}}$ is the fractional part of
the respective polarization charges in the plates of two capacitors: $n_{%
\mathrm{NR}}=\left\{ N_{\mathrm{NR}}\right\} $ and $n_{\mathrm{CPB}}=\left\{
N_{\mathrm{CPB}}\right\} $ with $N_{\mathrm{NR}}=C_{\mathrm{NR}}V_{\mathrm{NR%
}}/2e$ and $N_{\mathrm{CPB}}=C_{\mathrm{CPB}}V_{\mathrm{CPB}}/2e$.

Here we consider the Cooper-pair box formed by four capacitances, $C_{%
\mathrm{J}1}$, $C_{\mathrm{J}2}$, $C_{\mathrm{CPB}}$, and $C_{\mathrm{NR}}$ (%
$C_{\mathrm{J}}\gg C_{\mathrm{CPB}},C_{\mathrm{NR}}$). One of the plates of
the latter capacitor is formed by the NR, which is characterized by the
displacement at the midpoint $x$. This displacement is usually much smaller
than the distance $d$ between the plates, in which case the capacitance
between the NR and the qubit reads \cite{Lassagne09, Sillanpaa09, LaHaye09}

\begin{equation}
C_{\mathrm{NR}}(x)\approx C_{\mathrm{NR}0}+\left. \frac{\partial C_{\mathrm{%
NR}}}{\partial x}\right\vert _{0}x\equiv C_{\mathrm{NR}0}\left( 1+\frac{x}{%
\xi }\right) ,  \label{CNR(x)}
\end{equation}%
\begin{equation}
\xi ^{-1}=\frac{1}{C_{\mathrm{NR}0}}\left. \frac{\partial C_{\mathrm{NR}}}{%
\partial x}\right\vert _{0},\text{ \ }\xi \sim d\gg x\text{.}  \label{ksi}
\end{equation}%
(By the subscript $0$ here we mean the values at $x=0$; in what follows this
subscript is assumed). The displacement of the NR influences the qubit
through the changes in the polarization charge; to make this influence
significant, a large dc voltage $V_{\mathrm{NR}}$ (of the order of volts) is
applied. On the other side, the NR is biased by dc and rf voltages, $V_{%
\mathrm{GNR}}$ and $V_{\mathrm{RF}}$, through the capacitance $C_{\mathrm{GNR%
}}$, which provide its control and read-out

The influence of the qubit's dynamics on the nanomechanical resonator can be
described in different ways. In Appendix~A we present a detailed derivation
of the influence of the qubit's state through the voltage $V_{\mathrm{I}}$
and the average polarization charge $-2e\left\langle n\right\rangle $ of the
CPB on the NR's dynamics. An alternative, and maybe physically more
illustrative, approach is to describe the CPB as an effective capacitor,
which is the subject of Appendix~B. Here, in the main text, we present only
essential results, referring the interested reader to the Appendices.

As a result of the interaction between the qubit and the NR, the resonance
frequency of the NR is shifted (see Appendix~A). The result can be written
in the following form%
\begin{equation}
\frac{\Delta \omega _{\mathrm{NR}}}{\omega _{\mathrm{NR}}}=-\beta \frac{%
\partial \left\langle n\right\rangle }{\partial n_{\mathrm{g}}}=-\frac{\beta
}{2}\frac{\partial \left\langle \sigma _{z}\right\rangle }{\partial n_{%
\mathrm{g}}},  \label{DwNR_1}
\end{equation}%
\begin{equation}
\beta =\frac{1}{m\omega _{\mathrm{NR}}^{2}C_{\Sigma }}\left( \frac{C_{%
\mathrm{NR}}V_{\mathrm{NR}}}{\xi }\right) ^{2}.
\end{equation}%
The frequency shift $\Delta \omega _{\mathrm{NR}}$\ is defined by the \emph{%
derivative} of the average extra Cooper-pair number on the island $%
\left\langle n\right\rangle =0\cdot P_{0}+1\cdot P_{1}=P_{1}$. Here $P_{0}$ $%
(P_{1})$ stands for the probability of having $0$ $(1)$ extra Cooper pair.

Alternatively to the approach above, the effect of the qubit on the NR can
be described in terms of the effective (differential) capacitance, as
described in Appendix B, $C_{\mathrm{eff}}=\partial Q_{\mathrm{NR}}/\partial
V_{\mathrm{NR}}=C_{\mathrm{geom}}+C_{\mathrm{Q}}$, where the relevant \emph{%
quantum} capacitance is given by
\begin{equation}
C_{\mathrm{Q}}=\frac{C_{\mathrm{NR}}^{2}}{C_{\Sigma }}\frac{\partial
\left\langle n\right\rangle }{\partial n_{\mathrm{g}}}.  \label{CQ1}
\end{equation}%
The term \textquotedblleft quantum\textquotedblright\ capacitance is used
here to denote the (small) qubit-state-dependent addition to the classical
(geometric) capacitance. Obviously, Eq.~(\ref{DwNR_1}) can be rewritten in
terms of the quantum capacitance (\textit{cf}. discussion in Appendix~C for
the qubit-$LCR$ circuit system)%
\begin{equation}
\frac{\Delta \omega _{\mathrm{NR}}}{\omega _{\mathrm{NR}}}=-\widetilde{\beta
}\frac{C_{\mathrm{Q}}}{C_{\mathrm{NR}}},  \label{DwNR_2}
\end{equation}%
where $\widetilde{\beta }=\left( C_{\mathrm{\Sigma }}/C_{\mathrm{NR}}\right)
\beta $.

The qubit's density matrix in the energy representation (in the eigenbasis
of the time-independent Hamiltonian) can be parameterized in terms of the
respective Pauli matrices $\tau _{i}$, $\rho =\frac{1}{2}\left( X\tau
_{x}+Y\tau _{y}+Z\tau _{z}\right) $, as e.g. in Ref.~[%
\onlinecite{Shevchenko08}]. Here $Z=\left\langle \tau _{z}\right\rangle $ is
the difference between the occupation probabilities of the excited and
ground states. Now we express the probability of having one excess Cooper
pair, $P_{1}$, by changing from the energy basis to the charge basis, and
obtain
\begin{equation}
P_{1}=\frac{1}{2}\left( 1-\frac{\Delta }{\Delta E}X+\frac{\varepsilon _{0}}{%
\Delta E}Z\right) ,\text{ }\Delta E=\sqrt{\Delta ^{2}+\varepsilon _{0}^{2}}.
\end{equation}%
And this gives (after time-averaging over the driving period $2\pi /\omega $%
) for the quantum capacitance the following%
\begin{equation}
C_{\mathrm{Q}}\approx \frac{C_{\mathrm{NR}}^{2}}{C_{\Sigma }}\left( \frac{%
4E_{\mathrm{C}}\Delta ^{2}}{\Delta E^{3}}Z+\frac{\varepsilon _{0}}{2\Delta E}%
\frac{\partial Z}{\partial n_{\mathrm{g}}}\right) ,  \label{CQ2}
\end{equation}%
where we have taken into account that in the stationary state $X$ averages
to $0$.\cite{SAN}

As we can see from Eq.~(\ref{CQ2}), the quantum capacitance is defined by
the value $Z=\left\langle \tau _{z}\right\rangle $. In particular, we obtain
the quantum capacitance and the respective frequency shift in the
ground/excited ($\mathrm{g/e}$) state with $Z=\pm 1$%
\begin{equation}
\frac{\Delta \omega _{\mathrm{NR}}^{\mathrm{g/e}}}{\omega _{\mathrm{NR}}}%
=\mp \beta \frac{4E_{\mathrm{C}}\Delta ^{2}}{\Delta E^{3}}.  \label{CQ_ge}
\end{equation}%
This result, obtained in the semi-classical approach, is in agreement\emph{\
}with the one obtained in Ref.~\onlinecite{Irish03} and used in Ref.~%
\onlinecite{LaHaye09}. Equation~(\ref{CQ2}) is a more general result, where
the second term describes the sign-changing behavior near resonance. Namely,
when sweeping the gate voltage $n_{\mathrm{g}}$, the quantity $Z$ changes
from $-1$, far from resonance (in the ground state), to $0$ in resonance
(when the levels are equally populated). This describes the maximum of $Z$
in resonance and the change of its derivative $\partial Z/\partial n_{%
\mathrm{g}}$ from positive, in the left vicinity of the resonance, to
negative, to the right of the resonance point. Thus, the resulting behavior
of the observable (either $\Delta \omega _{\mathrm{NR}}$ or $C_{\mathrm{Q}}$%
) is defined by the competition of the two terms in Eq.~(\ref{CQ2}). In what
follows we will use Eq.~(\ref{CQ2}) for the superposition states (which
appear under driving).\cite{Sillanpaa06} Note that a similar approach for
calculating the effective (quantum) inductance was used in Refs.~%
\onlinecite{Shnyrkov06,
Shevchenko08}.

The dissipative dynamics can be described with the Bloch equations written
in the energy representation (where relaxation appears naturally).\ To
characterize dissipation we use a result of the spin-boson model with the
spectral density defined with the dimensionless parameter $\alpha $, $%
J(\omega )=\alpha \hbar \omega $, see e.g. Ref.~\onlinecite{Temch11} and
references therein, while the low-frequency $1/f$ noise is described by the
peak of $J(\omega )$ at $\omega \approx 0$. Then the relaxation and
dephasing times are defined by the spectral density at $\omega \approx
\Delta E$ and $\omega \approx 0$ respectively as following%
\begin{equation}
T_{1}^{-1}=\alpha \frac{\Delta ^{2}}{2\hbar \Delta E}\coth \frac{\Delta E}{%
2k_{B}T},  \label{T1}
\end{equation}%
\begin{equation}
T_{2}^{-1}=\frac{1}{2}T_{1}^{-1}+\frac{k_{B}T}{\hbar }\frac{\varepsilon
_{0}^{2}}{\Delta E^{2}}(\alpha +\frac{B}{2\pi })\approx B\frac{k_{B}T}{h}%
\frac{\varepsilon _{0}^{2}}{\Delta E^{2}}.  \label{T2}
\end{equation}%
Here the (relatively large) phenomenological parameter $B$ was introduced to
describe the low-frequency $1/f$ noise. We note that alternatively the
low-frequency noise could be taken into account as the averaging of the
final solution resulting in some blurring of the resonances, as e.g. in Ref.~%
\onlinecite{Sillanpaa06}. The values for the relaxation and dephasing times
define the shape of the resonances (as for example it is later described by
Eqs.~(\ref{Pup}, \ref{De})). In this way, the width of the resonances can be
used for the estimation of the dephasing rate. In our case, we have taken $%
\alpha $ and $B$ as the fitting parameters, to obtain better resemblance
with the experimental results.

\begin{figure}[t]
\includegraphics[width=8cm]{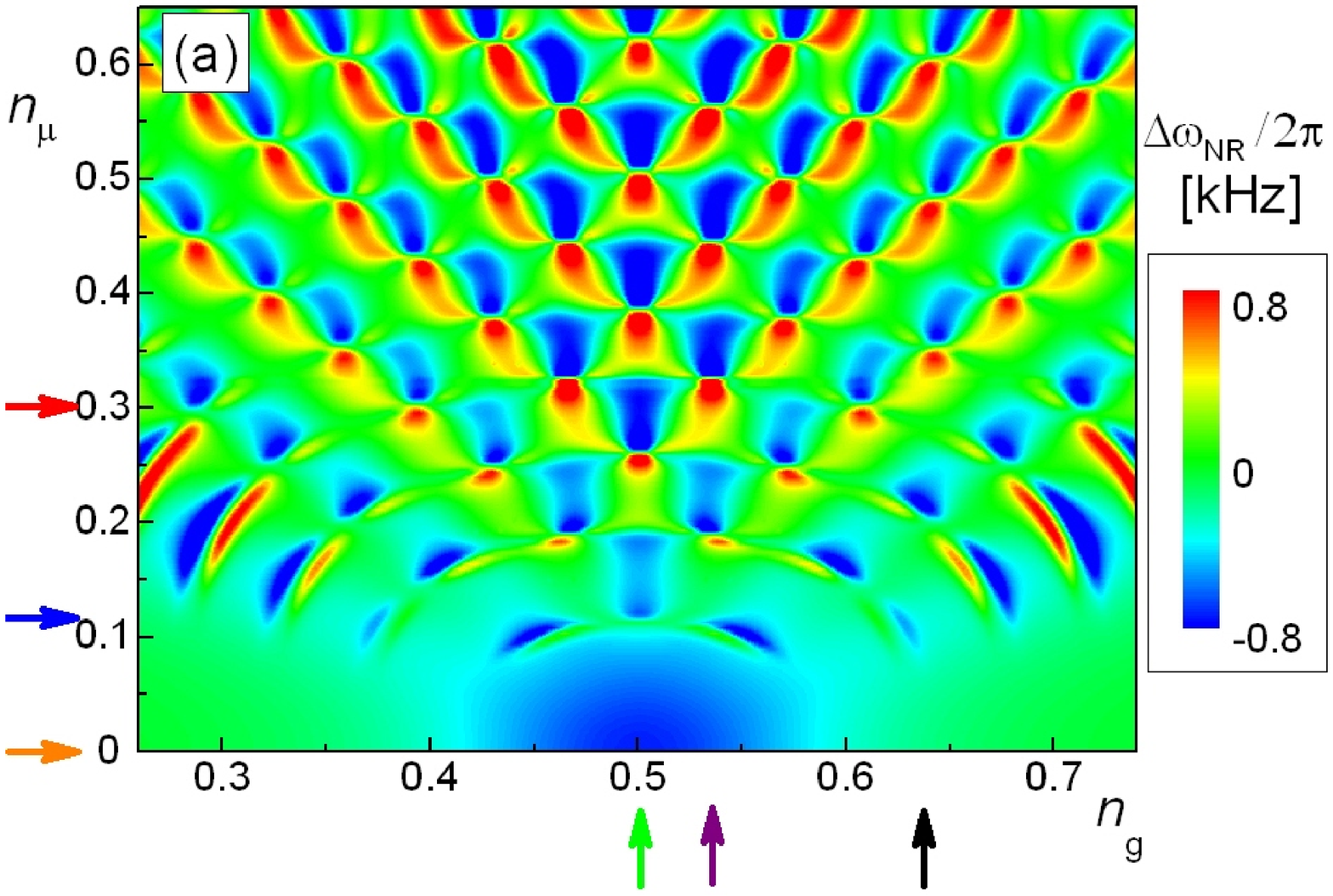} \includegraphics[width=8cm]{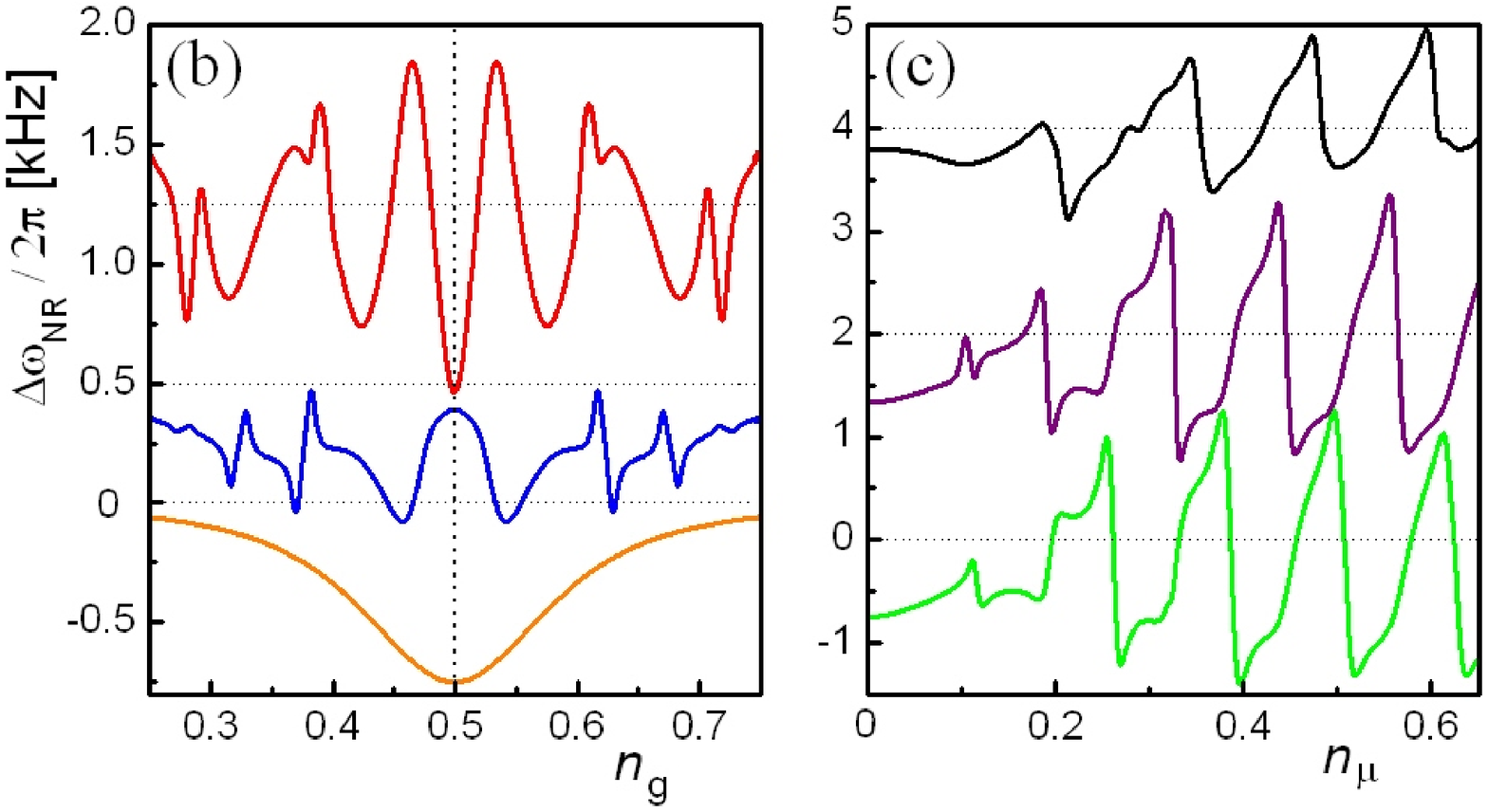}
\caption{(Color online) LZS interferometry probed via the resonator's
frequency shift $\Delta \protect\omega _{\mathrm{NR}}$. (a) The frequency
shift versus the energy bias ($n_{\mathrm{g}}$) and the driving amplitude ($%
n_{\protect\mu }$). Arrows show the values of $n_{\protect\mu }$ and $n_{%
\mathrm{g}}$ at which the graphs (b) and (c) are plotted as functions of $n_{%
\mathrm{g}}$ and $n_{\protect\mu }$, respectively. The upper curves were
shifted for clarity. The parameters for calculations were taken close to the
ones of Ref.~[\onlinecite{LaHaye09}]: $\protect\omega _{\mathrm{NR}}/2%
\protect\pi =58$ MHz, $E_{\mathrm{J}0}/h=13$ GHz, $E_{\mathrm{C}}/h=14$ GHz,
$\protect\omega /2\protect\pi =4$ GHz, $k_{B}T/h=2$ GHz, $\protect\alpha %
=0.005$, $B=0.2$, and the proportionality coefficient $\protect\beta $
defined by the qubit-NR coupling constant $\protect\lambda $ from %
Ref.~[\onlinecite{LaHaye09}]: $\hbar \protect\lambda ^{2}/\protect\pi E_{\mathrm{J}%
0}=\protect\beta \cdot E_{\mathrm{C}}\protect\omega _{\mathrm{NR}}/\protect%
\pi E_{\mathrm{J}0}=1.6$\ kHz.}
\label{Fig:Dw}
\end{figure}

We display the direct LZS interferometry in Fig.~\ref{Fig:Dw}, where the
resonator's frequency shift $\Delta \omega _{\mathrm{NR}}$ was calculated
with Eqs.~(\ref{DwNR_2}) and (\ref{CQ2}). Figure~\ref{Fig:Dw} demonstrates
that our formalism is valid for a description of the experimentally
measurable quantities: the quantum capacitance or the resonant frequency
shift \cite{LaHaye09}$^{,}$ \cite{Sillanpaa06} (see also Appendix~C). Such a
description allows to correctly find the position of the resonance peaks in
the interferogram and to demonstrate the sign-changing behavior of the
quantum capacitance, which relates to the measurable quantities. The
appearance of the interferogram depends on several factors: the values of
the qubit parameters, the model for the dissipative environment (such as
Eqs.~(\ref{T1},~\ref{T2}) and the parameters $\alpha $ and $B$), the value
of the bias current (which distorts the shape of the resonances, as
demonstrated in Ref.~[\onlinecite{Shevchenko08}]). Moreover, the formalism
presented above is valid for the case where the qubit's dynamics is much
faster than the NR's dynamics; otherwise one should study the cooperative
dynamics of the composite system; see, e.g., discussions in Refs.~[%
\onlinecite{Sillanpaa06}] and [\onlinecite{Shevchenko08}]. However, we will
not go here into more detailed calculations, since our aim was to
demonstrate the simplest approach for the description of the experiment in
Ref.~[\onlinecite{LaHaye09}].

\section{The bias influenced by the resonator: Problem for the inverse
interferometry}

Let us now consider the qubit's bias $\varepsilon _{0}$, Eq.~(\ref{e0}), as
a function of the NR's displacement $x$. For small $x\ll \xi $, we have the
expansion (\ref{CNR(x)}), which results in the decomposition of the bias%
\begin{equation}
\varepsilon _{0}(x)\approx \varepsilon _{0}^{\ast }(n_{\mathrm{g}})+\delta
\varepsilon _{0}(x),  \label{e0(x)2}
\end{equation}%
where%
\begin{eqnarray}
\varepsilon _{0}^{\ast }(n_{\mathrm{g}}) &=&8E_{\mathrm{C}}\left( n_{\mathrm{%
g}}-1/2\right) , \\
\delta \varepsilon _{0}(x) &=&8E_{\mathrm{C}}\,n_{\mathrm{NR}}\frac{x}{\xi }.
\end{eqnarray}%
Here we have used the fact that $x\ll \xi $ and $C_{\mathrm{NR}}\ll
C_{\Sigma }$.

The Hamiltonian of the qubit (\ref{H(t)}) with the parameter-dependent bias $%
\varepsilon _{0}(x)$ brings us to the following problem. Let us assume that
the qubit's state is known (i.e., this is measured by a device whose details
we do not consider here for simplicity; see Refs.~[%
\onlinecite{Sillanpaa05,
Duty05, Shnyrkov06, Oliver05}] for different realizations of the ways to
probe the qubit's state). Given the known qubit state, we aim to find the
Hamiltonian's parameters. Particularly, we are interested in the
parameter-dependent bias $\varepsilon _{0}(x)$.

On one hand, we can study here the general (``reverse engineering") problem
in the spirit of Refs.~[\onlinecite{Garanin02, Berry09}]. On the other hand,
we aim to provide the basis for measuring the NR's position $x$ by means of
probing the qubit's state, while $x=x(t)$ is considered a slow
time-dependent function.

In what follows we will consider the driven qubit's state with emphasis on
finding optimal driving and controlled offset parameters ($A$, $\omega $,
and $\varepsilon _{0}^{\ast }$) for the resolution of the small bias
component $\delta \varepsilon _{0}$. We will assume that the dynamics of the
parameter $x$ is slow enough not to be considered during the measurement
process. Depending on this slowness, the measurement might have to involve
only one passage of the avoided crossing, or it can involve long-time
driving and stationary-state equilibrium of the qubit. Our aim is to find a
sensitive probe for small $\delta \varepsilon _{0}$. For high sensitivity we
require substantial changes in the qubit's state for small changes of $%
\varepsilon _{0}$\ given by $\delta \varepsilon _{0}$. For a quantitative
definition of the sensitivity one can consider the derivative of the
probability with respect to the bias $\varepsilon _{0}$.

\section{Results for the inverse LZS interferometry}

In this section we consider the inverse problem for the qubit's dynamics, in
particular how to infer the qubit's bias $\varepsilon _{0}$ from the
measured qubit state. For concreteness, we consider the qubit driven by the
bias $\varepsilon (t)=\varepsilon _{0}+A\sin \omega t$. For purposes of
analyzing the short-time dynamics, one would consider a single passage or a
sequence of a small number of passages through the avoided level crossing.
If the time-dependence of the bias $\varepsilon _{0}(x)$ is so slow that the
multiple-passage dynamics is relevant, then the stationary qubit state can
be considered.

\subsection{Single passage: non-linearity in the Landau-Zener problem}

\begin{figure}[t]
\includegraphics[width=8.5cm]{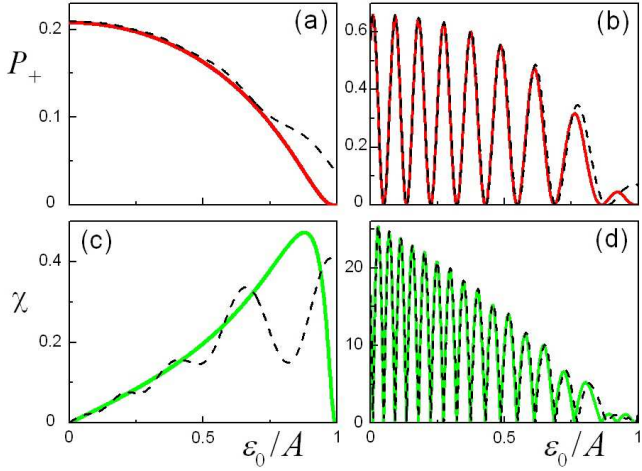}
\caption{(Color online) Upper-level excitation probability $P_{+}$ after (a)
single passage and (b) double passage, plotted for $A/\Delta =5$ and $\hbar
\protect\omega /\Delta =0.2$, versus the bias $\protect\varepsilon _{0}$.
The sensitivity to the changes of the bias $\protect\varepsilon _{0}$,
defined as the derivative, $\protect\chi =\left\vert dP_{+}/d(\protect%
\varepsilon _{0}/A)\right\vert $, is plotted in (c) and (d), respectively.
Solid lines were plotted with Eqs. (\protect\ref{PLZ}) and (\protect\ref{Pp2}%
), while dashed lines were calculated numerically.}
\label{PIPII}
\end{figure}

The linearization of the bias in the vicinity of the avoided crossing (where
$\varepsilon (t)=0$) results in the approximation that this region is swept
at the $\varepsilon _{0}$-dependent rate $A\omega \sqrt{1-(\varepsilon
_{0}/A)^{2}}$ (for details see Ref.~[\onlinecite{SAN}]). The respective
probability of the non-adiabatic transition to the upper adiabatic level is
given by the Landau-Zener formula%
\begin{equation}
P_{+}^{(I)}=P_{\mathrm{LZ}}=\exp \left( -\frac{\gamma }{\sqrt{1-(\varepsilon
_{0}/A)^{2}}}\right) ,\text{ }\gamma =\frac{\pi }{2}\frac{\Delta ^{2}}{%
A\hbar \omega }.  \label{PLZ}
\end{equation}%
In other words, the non-linear dependence of the bias on time has the effect
that the Landau-Zener probability depends on $\varepsilon _{0}$ (see also
Ref.~[\onlinecite{Garanin02b}]), which is demonstrated in Fig.~\ref{PIPII}%
(a). We note that here $\left\vert \varepsilon _{0}\right\vert <A$ and the
formula~(\ref{PLZ}) gives numerically incorrect results when $\varepsilon
_{0}$ tends to $A$.

To quantify the sensitivity of the transition probability to small changes
in the bias, in Fig.~\ref{PIPII}(c) we plot the derivative of the excitation
probability $P_{+}^{(I)}$ with respect to $\varepsilon _{0}$. We can see
that the non-linearity of the bias results in an increase of the sensitivity.

For the single-passage case it is straightforward, from Eq.~(\ref{PLZ}), to
find the solution for the inverse problem $\varepsilon _{0}=\varepsilon
_{0}(P_{+}^{(I)})$. In particular, in the case $\varepsilon _{0}^{\ast }=0$
and $\delta \varepsilon _{0}\ll A$ we have%
\begin{equation}
P_{\mathrm{LZ}}\approx P_{\mathrm{LZ},0}\left[ 1-\frac{\gamma }{2}\left(
\frac{\delta \varepsilon _{0}}{A}\right) ^{2}\right] ,\text{ }P_{\mathrm{LZ}%
,0}=e^{-\gamma },  \label{PLZ0}
\end{equation}%
and the solution for the inverse problem becomes%
\begin{equation}
\frac{\delta \varepsilon _{0}}{A}=\sqrt{\frac{2}{\gamma }\left( 1-\frac{P_{%
\mathrm{LZ}}}{P_{\mathrm{LZ},0}}\right) }.
\end{equation}

\subsection{Double passage: St\"{u}ckelberg oscillations}

Next, consider the situation where the avoided crossing region is passed
twice. For example, the qubit can be driven by a sinusoidal pulse of length $%
2\pi /\omega $. Alternatively, triangular pulses can be used to drive the
qubit twice through the avoided-level crossing, as in Refs.~[%
\onlinecite{Sun10,Huang11}]. In both cases, the double-passage process can
make use of quantum interference to increase the sensitivity of our problem
through the accumulation of the St\"{u}ckelberg phase. \cite{Suzuki10}

The upper-level excitation probability after the double-passage is \cite{SAN}%
\begin{equation}
P_{+}^{(II)}=4P_{\mathrm{LZ}}(1-P_{\mathrm{LZ}})\sin ^{2}(\zeta _{2}+\varphi
_{\mathrm{S}}),  \label{Pp2}
\end{equation}%
where $\zeta _{2}$ is the phase acquired during the evolution between
anticrossings at $t_{2}$ and $t_{1}+2\pi /\omega $:%
\begin{equation}
\zeta _{2}=\frac{1}{2\hbar }\int_{t_{2}}^{t_{1}+2\pi /\omega }\sqrt{\Delta
^{2}+\varepsilon (t)^{2}} \; dt,  \label{dzeta2}
\end{equation}%
and $\varphi _{\mathrm{S}}$ is the Stokes phase.

St\"{u}ckelberg oscillations, described by Eq.~(\ref{Pp2}), are demonstrated
in Fig.~\ref{PIPII}(b) for $0<\varepsilon _{0}/A<1$. The respective
sensitivity is shown in Fig.~\ref{PIPII}(d). The agreement of the analytical
formulas and numerical calculations is remarkable (as demonstrated in Fig.~%
\ref{PIPII}). One can notice that the sharper the St\"{u}ckelberg
oscillations, the higher the sensitivity. This is related to the period of
the St\"{u}ckelberg oscillations, which decreases with increasing $A/\omega $%
. Here we also note that $P_{+}^{(II)}(\varepsilon _{0})$ is not a symmetric
function, and the period of the St\"{u}ckelberg oscillations is smaller for $%
\varepsilon _{0}<0$ than for $\varepsilon _{0}>0$. Therefore, using negative
values of $\varepsilon _{0}$ results in slightly higher sensitivity than
what is shown in Fig.~\ref{PIPII}(d).

The factor $P_{\mathrm{LZ}}(1-P_{\mathrm{LZ}})$\ in Eq.~(\ref{Pp2}) is
described by the one-passage problem above. Consider the term $\cos
^{2}\zeta _{2}$. For $\varepsilon _{0}^{\ast }=0$ and $\delta \varepsilon
_{0}\ll A$ we have\cite{SAN} $\zeta _{2}\approx \frac{A}{\hbar \omega }-%
\frac{\pi }{2}\frac{\delta \varepsilon _{0}}{\hbar \omega }$. For example,
for $\frac{A}{\hbar \omega }=2k\pi +\frac{\pi }{4}$ we obtain%
\begin{equation}
P_{+}^{(II)}\approx 2P_{\mathrm{LZ}}(1-P_{\mathrm{LZ}})\left( 1+\pi \frac{%
\delta \varepsilon _{0}}{\hbar \omega }\right) .  \label{PII}
\end{equation}%
This describes a linear dependence on the small bias $\delta \varepsilon
_{0} $, which is a significant increase in sensitivity as compared to the
quadratic dependence on $\delta \varepsilon _{0}$ in the single-passage case
above, Eq.~(\ref{PLZ0}). If the decoherence is negligibly small, one can
further increase the sensitivity of the excitation probability to small
changes in the bias due to interference by considering multiple-passage case.

The formula (\ref{PII}) can be conveniently used to make quantitative
estimates. Consider this for the example of the qubit-nanomechanical
resonator system as in Ref.~[\onlinecite{LaHaye09}]. First, to increase the
sensitivity of the changes of $P_{+}^{(II)}$ with respect to $\delta
\varepsilon _{0}$, we choose the smallest possible frequency $\omega $. In
our case the driving period should exceed the decoherence time $T_{2}$ and
the NR oscillation period $2\pi /\omega _{\mathrm{NR}}$. For superconducting
qubits $T_{2}$ is typically higher than $1$ $\mu $s. Then, we are limited by
the relation $\omega >\omega _{\mathrm{NR}}$, and we take $\omega /2\pi \sim
0.1$ GHz. We choose the parameters $A(n_{\mu })$ and $\Delta (\Phi )$ such
that\ $P_{\mathrm{LZ}}\sim 1/2$. Assuming $n_{NR}=1$ and $8E_{C}/h=100$ GHz,
we obtain the change of the probability with changes in the NR's
displacement $\Delta P_{+}^{(II)}=10^{3}x/\xi $. This means that for probing
a displacement of $x\sim 10^{-5}\xi $, one has to be able to measure
population changes $P_{+}^{(II)}\sim 0.01$. This level of accuracy is
achievable with superconducting qubits.\cite{Lucero08}

\subsection{Multiple passage: stationary solution}

Now we assume that what is relevant for our inverse problem is the
stationary state of the driven qubit. To analyze the analytical expressions,
we consider two limiting cases.

\subsubsection{\textbf{Slow-passage limit}}

\begin{widetext}

\begin{figure}[t]
\includegraphics[width=11cm]{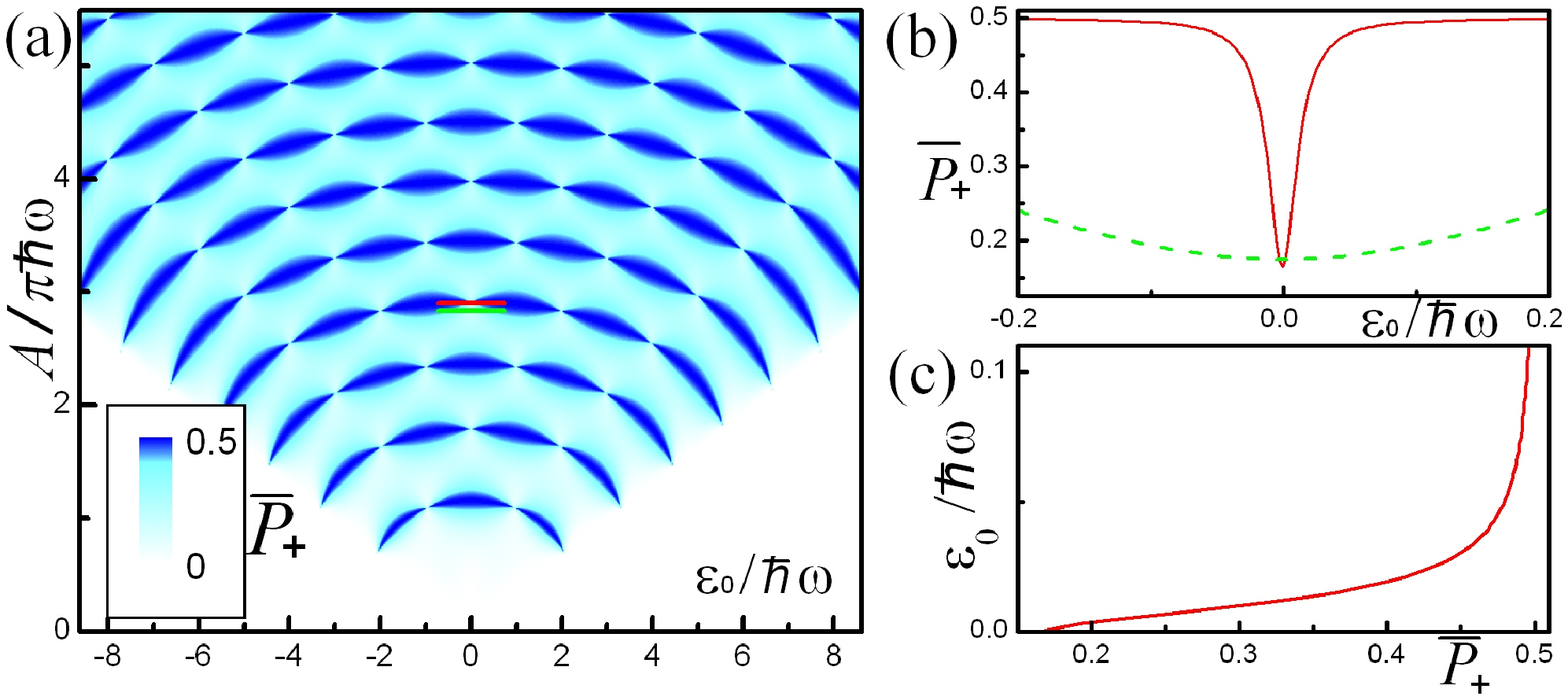} %
\includegraphics[width=11cm]{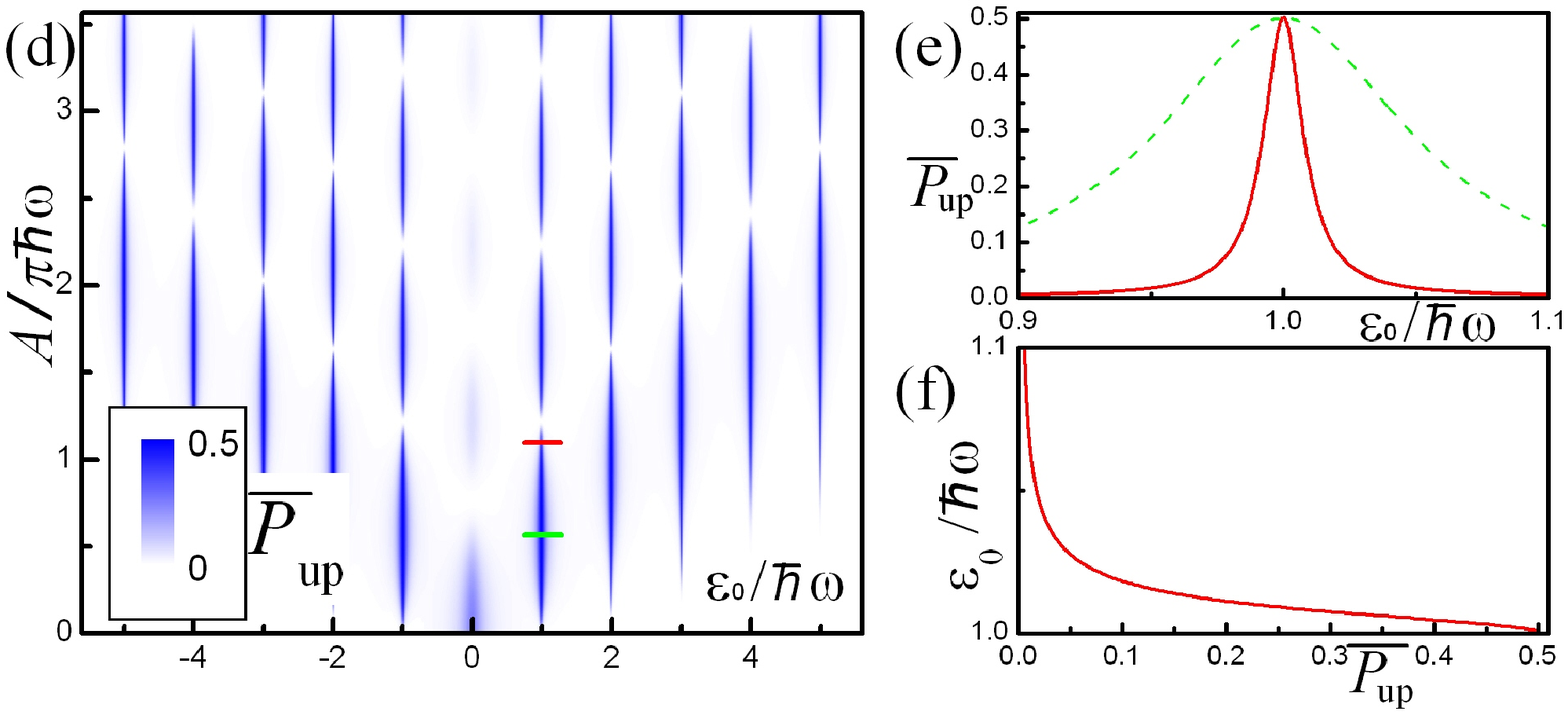}
\caption{(Color online) Slow-passage and fast-passage LZS interferometry of a
qubit. (a) and (d): the time-averaged upper-level occupation
probabilities, defined in the adiabatic ($\overline{P_{+}}$) and
diabatic ($\overline{P}_{\mathrm{up}}$) bases, as functions of the
bias $\varepsilon _{0}$ and driving amplitude $A$. The parameters
are the same as for Fig. \protect\ref{Fig:Dw} except for the
frequency: (a) $\protect\omega
/2\protect\pi =6.5$ GHz $<\Delta /h$ and (d) $\protect\omega /2\protect\pi =20$ GHz $%
>\Delta /h$. (b) and (e): Cross-sections for the respective dependencies of the
upper-level occupation probabilities as functions of the bias
along the horizontal dashes shown in red and green in (a) and (d).
(c) and (f): Inverse graphs, which show the dependence of the bias
on the upper-level occupation probabilities (assuming that
$\varepsilon _{0}$ lies on the right-hand side of the resonance
peak).} \label{Fig2}
\end{figure}

\end{widetext}

For the analytical description of the upper-level occupation probability in
the adiabatic limit, when $\gamma >1$, we use the following formula from
Ref.~[\onlinecite{SAN}]

\begin{equation}
\overline{P_{+}}=\frac{P_{\mathrm{LZ}}(1-\cos \zeta _{+}^{\prime }\cos \zeta
_{-})}{\sin ^{2}\zeta _{+}^{\prime }+2P_{\mathrm{LZ}}(1-\cos \zeta
_{+}^{\prime }\cos \zeta _{-})},  \label{Pp_with_offset}
\end{equation}%
where
\begin{eqnarray}
\zeta _{+}^{\prime } &=&\zeta _{1}+\zeta _{2},\text{ }\zeta _{-}=\zeta
_{1}-\zeta _{2},\text{ }  \notag \\
\zeta _{1} &=&\frac{1}{2\hbar }\int\limits_{t_{1}}^{t_{2}}\sqrt{\Delta
^{2}+\varepsilon (t)^{2}}dt,
\end{eqnarray}%
and $\zeta _{2}$ is given by Eq.~(\ref{dzeta2}). Formula~(\ref%
{Pp_with_offset}) is illustrated in Fig.~\ref{Fig2}(a). Consider $%
\varepsilon _{0}^{\ast }=0$, then for strong driving, $A\gg \Delta $, we
have
\begin{equation}
\zeta _{-}\approx \frac{\pi \delta \varepsilon _{0}}{\hbar \omega },\text{ \
}\zeta _{+}^{\prime }\approx \frac{2A}{\hbar \omega }-\frac{\delta
\varepsilon _{0}^{2}}{A\hbar \omega }.
\end{equation}%
Analyzing the interferogram in Fig.~\ref{Fig2}(a), we find the possibility
to obtain a sensitive working point with a driving amplitude a little bit
lower than the one where the width of the resonance line tends to zero, that
is $2A/\hbar \omega =2\pi n-a$, $a\ll 1$ [see the red and green dashes in
Fig.~\ref{Fig2}(a)]. It follows that%
\begin{equation}
\overline{P_{+}}\approx \frac{1}{2}\frac{P_{\mathrm{LZ}}\left( \pi \delta
\varepsilon _{0}/\hbar \omega \right) ^{2}}{a^{2}+P_{\mathrm{LZ}}\left( \pi
\delta \varepsilon _{0}/\hbar \omega \right) ^{2}},
\end{equation}%
which is equal to zero at $\delta \varepsilon _{0}=0$ and quickly tends to $%
1/2$ with increasing $\delta \varepsilon _{0}$. This is demonstrated in Fig.~%
\ref{Fig2}(b).

\subsubsection{\textbf{Fast-passage limit}}

In the fast-passage and strong-driving regime (where $\gamma \ll 1$), the
rotating-wave approximation gives for the upper-level occupation probability
\cite{Oliver05, Ashhab07}%
\begin{eqnarray}
\overline{P}_{\mathrm{up}} &=&\frac{1}{2}\sum_{k}\frac{\Delta _{k}^{2}}{%
\frac{\hbar ^{2}}{T_{1}T_{2}}+\frac{T_{2}}{T_{1}}(\varepsilon _{0}-k\hbar
\omega )^{2}+\Delta _{k}^{2}},\text{ }  \label{Pup} \\
\Delta _{k} &=&\Delta J_{k}\left( A/\hbar \omega \right) ,
\end{eqnarray}%
where $J_{k}$\ is the Bessel function. Formula~(\ref{Pup}) is demonstrated
in Fig.~\ref{Fig2}(d). If the relaxation is not taken into account, then in
the vicinity of the $k$-th resonance (where $\varepsilon _{0}^{\ast }=k\hbar
\omega $) we obtain the Lorentzian dependence on the small bias shift $%
\delta \varepsilon _{0}$:%
\begin{equation}
\overline{P}_{\mathrm{up}}=\frac{1}{2}\frac{\Delta _{k}^{2}}{\delta
\varepsilon _{0}^{2}+\Delta _{k}^{2}}.
\end{equation}%
This describes the resonance peak, $\overline{P}_{\mathrm{up}}=1/2$, at $%
\delta \varepsilon _{0}=0$, which is demonstrated graphically in Fig.~\ref%
{Fig2}(e). Its width is defined by $\Delta _{k}$ and is minimized for the
values of $A/\hbar \omega $ in the vicinity of the zeros of the Bessel
function. With relaxation taken into account, the sensitivity is defined by
the half-width of the resonances, given by
\begin{equation}
\Delta \varepsilon _{0}^{(k)}=\frac{\sqrt{T_{1}T_{2}\Delta _{k}^{2}+\hbar
^{2}}}{T_{2}}.  \label{De}
\end{equation}%
This means that to increase the sensitivity, which is related to the
sharpness of the resonances, one has to decrease the decoherence rate.

Here we note that it was assumed that the measurement time is much smaller
than the resonator's period, $T_{\mathrm{meas}}\ll 2\pi /\omega _{\mathrm{NR}%
}$. On the other hand, to reach a stationary state, the measurement time
should be larger than the relaxation time, $T_{1,2}<T_{\mathrm{meas}}$. This
means that the results presented in this subsection are relevant for qubits
with short relaxation times and for resonators with small frequencies.
Alternatively, one should solve the problem which explicitly takes into
account $x=x(t)$.

Formula~(\ref{De}) allows us to make estimates, as we did at the end of the
previous subsection. For $A/\hbar \omega $ equal to one of the
Bessel-function zeros and for $T_{2}=4\mathrm{ns}\ll 2\pi /\omega _{\mathrm{%
NR}}$, we obtain that the probability $\overline{P}_{\mathrm{up}}$\ changes
by about $1/4$ when the bias changes by $\Delta \varepsilon _{0}/h\sim 0.25$
GHz. On the other hand, we have seen that $\delta \varepsilon _{0}/h\sim
100(x/\xi )$ GHz. This means that in order to observe changes $x\sim
10^{-5}\xi $, one has to distinguish changes in $\overline{P}_{\mathrm{up}}$
$\sim 10^{-3}$, which is also possible, in principle.\cite{Lucero08}

\subsection{Inverse interferometry: qubit probes resonator}

The idea of the measurement procedure, presented in Fig.~\ref{Fig2}, could
be as follows. Driving the qubit in a wide range of parameters is done first
to plot the interferogram as in Fig.~\ref{Fig2}(a) and/or (d). Then a region
of high sensitivity, where small changes in the qubit bias result in large
changes in the final state, is chosen. Examples of such high-sensitivity
regions are shown in Fig.~\ref{Fig2}(b) and/or (e).

From Fig.~\ref{Fig2} we can see that both the slow-passage limit,
demonstrated in Fig.~\ref{Fig2}(a-c), and the fast passage limit [Fig.~\ref%
{Fig2}(d-f)] can be used for the solution of the inverse problem. The choice
of the optimal working point and its vicinity will depend on the specific
parameters of the problem. For illustration, in Fig.~\ref{Fig2}(a) and (d)
we marked by red and green small dashes two possibilities of having the\ dip
in Fig.~\ref{Fig2}(b) or the peak in Fig.~\ref{Fig2}(e) being narrow (red
curves) or relatively wide (green curves).

In principle, a low-amplitude slice near the bottom of Fig.~\ref{Fig2}(d)
can be used to obtain a sharp resonance peak, as in Fig.~\ref{Fig2}(e).
However, based on the results of Refs.~[\onlinecite{SAN, Du10}], it seems
that the width of the resonances might be increased more for low-amplitude
driving due to the influence of the noise and decoherence. From the
experimental point of view the best strategy is probably to obtain a wide
range interferogram and then choose a narrow resonance.

One can now bias the qubit at a high-sensitivity point, apply a
\textquotedblleft measurement pulse\textquotedblright\ to the qubit, measure
its state at the end of the pulse and extract the resonator's position $x$
from the measured qubit's state, see Fig.~\ref{Fig2}(c) and (f), where $%
\varepsilon _{0}$ (which parametrically depends on $x$) is plotted as a
function of the qubit's occupation probability.

It should be noted here that the measurement pulse, which is essentially a
driving signal applied to the qubit, can take a short duration at the
beginning of the measurement process. Afterwards the final state of the
qubit is read out in the absence of any driving fields. As a result, issues
that only affect the qubit on relatively large timescales, e.g. dephasing
and the slow measurement of the qubit's state, do not affect the qubit's
ability to accurately measure the instantaneous position of the resonator.
It should also be noted that this measurement procedure is a single-shot
type of measurement and not a continuous measurement. One could in principle
use several qubits in order to perform multiple measurements on the state of
the resonator.

\section{Conclusions}

We have analyzed a measurement scheme where a qubit is probed via a quantum
capacitance. We demonstrated the sign-changing behavior of the quantum
capacitance where the strongly-driven qubit exhibits a LZS interferogram.
Our semi-classical calculations were used to describe recent experimental
results\cite{LaHaye09} for the LZS interferometry of the qubit probed by a
NR.

Then, motivated by the experimental work by LaHaye et al.~[%
\onlinecite{LaHaye09}], we formulated the inverse problem. The inverse LZS
problem was formulated and solved for a generic two-level system in several
driving regimes. More specifically, we have split the quasi-constant bias $%
\varepsilon _{0}$ into an externally-controlled part $\varepsilon _{0}^{\ast
}(n_{\mathrm{g}})$ and a small part $\delta \varepsilon _{0}(x)$ that is to
be measured through the qubit's state. For the qubit-NR system the former
can be changed through the gate voltage to realize the most efficient
measurement working point; the latter was assumed to be a function of the
NR's displacement $x$.

We have shown how the inverse problem can be used for defining the NR's
displacement. First, one should find (measure) the direct LZS interferogram
(in a wide range of parameters). This allows finding the qubit's parameters
and choosing the optimal bias $\varepsilon _{0}^{\ast }$. Then, fixing the
qubit's parameters at the optimal working point, small changes due to the
slow NR's motion may be used for measuring its displacement.

\begin{acknowledgments}
SNS thanks V.I. Shnyrkov, O.G. Turutanov, and A.M. Zagoskin for useful
discussions. SNS was partly supported by the NAS of Ukraine (Project No.
04/10-N) and DKNII (Project No. M/411-2011). SA and FN were partially
supported by the Laboratory for Physical Science, National Security Agency,
Army Research Office, NSF grant No. 0726909, JSPS-RFBR contract No.
09-02-92114, Grant-in-Aid for Scientific Research (S), MEXT Kakenhi on
Quantum Cybernetics, and the JSPS-FIRST program.
\end{acknowledgments}

\appendix

\section{Semi-classical theory for the qubit-resonator system}

In this Appendix we consider the semi-classical theory for the qubit-NR
system. The equation for the displacement $x$ of the classical NR with
effective mass $m$, quality factor $Q$, eigenfrequency $\omega _{0}$ and
driven by the external force $F$, is
\begin{equation}
m\frac{d^{2}x}{dt^{2}}+\frac{m\omega _{0}}{Q}\frac{dx}{dt}+m\omega
_{0}^{2}x=F.  \label{eq4x}
\end{equation}%
In our problem, presented in Fig.~\ref{Fig:scheme}, the NR is influenced by
the voltage difference from both sides. On one side (to the right of the NR
in Fig.~\ref{Fig:scheme}) the voltage difference contains the large constant
part, $\Delta V=V_{\mathrm{NR}}-V_{\mathrm{GNR}}$, and the small rf driving
component, $V_{\mathrm{RF}}=V_{\mathrm{A}}\cos \omega _{\mathrm{rf}}t$. The
force due to these voltages is
\begin{eqnarray}
F_{\mathrm{GNR}} &=&\frac{1}{2}\frac{\partial }{\partial x}\left[ C_{\mathrm{%
GNR}}(V_{\mathrm{NR}}-V_{\mathrm{GNR}}-V_{\mathrm{RF}})^{2}\right]  \notag \\
&\approx &\frac{1}{2}\left( \frac{\partial C_{\mathrm{GNR}}}{\partial x}%
\right) \Delta V^{2}-F_{\mathrm{A}}\cos \omega _{\mathrm{rf}}t,
\label{F_GNR}
\end{eqnarray}%
where $F_{\mathrm{A}}=\left( \partial C_{\mathrm{GNR}}/\partial x\right)
\cdot \Delta V\cdot V_{\mathrm{A}}$. From the other side (left side of the
NR in Fig.~\ref{Fig:scheme}) the voltage difference is defined by the
island's voltage $V_{\mathrm{I}}$. The respective force is%
\begin{eqnarray}
F_{\mathrm{NR}} &=&\frac{1}{2}\frac{\partial }{\partial x}\left[ C_{\mathrm{%
NR}}(V_{\mathrm{NR}}-V_{\mathrm{I}})^{2}\right]  \notag \\
&\approx &\frac{1}{2}\left( \frac{\partial C_{\mathrm{NR}}}{\partial x}%
\right) V_{\mathrm{NR}}^{2}-V_{\mathrm{NR}}\frac{\partial }{\partial x}%
\left( C_{\mathrm{NR}}V_{\mathrm{I}}\right) .  \label{F_NR}
\end{eqnarray}

In the Coulomb-blockade regime, the voltage $V_{\mathrm{I}}$ is defined by
the quantum-mechanically averaged island's charge $-2en$, which is given by
the sum of the charges on the plates of the capacitors that define the
island,%
\begin{equation}
-2en=Q_{\mathrm{J1}}+Q_{\mathrm{J2}}-Q_{\mathrm{CPB}}-Q_{\mathrm{NR}}.
\end{equation}%
For the island's voltage it follows that%
\begin{equation}
V_{\mathrm{I}}=\frac{2e(N_{\mathrm{g}}+n_{\mu }\sin \omega t-n)}{C_{\Sigma }}%
,
\end{equation}%
\begin{equation}
N_{\mathrm{g}}=\frac{C_{\mathrm{NR}}V_{\mathrm{NR}}}{2e}+\frac{C_{\mathrm{CPB%
}}V_{\mathrm{CPB}}}{2e}\equiv N_{\mathrm{NR}}+N_{\mathrm{CPB}}.
\end{equation}

Here we note that to obtain the charging Hamiltonian of the CPB in the
two-state approximation, we consider $N_{\mathrm{g}}=N+n_{\mathrm{g}}$ close
to a half-integer number, where $N$ is the integer part of $N_{\mathrm{g}}$,
and $n_{\mathrm{g}}=\left\{ N_{\mathrm{g}}\right\} $ is the fractional part.
Then, with $n=N+\widehat{n}$ and $n_{\mu }<1$, we obtain for $H_{\mathrm{CPB}%
}=C_{\Sigma }V_{\mathrm{I}}^{2}/2$ the charging part of Hamiltonian, Eq.~(%
\ref{H(t)}). Here the operator for the extra Cooper-pair number $\widehat{n}%
=(1+\sigma _{z})/2$ acts on the \textquotedblleft charge" basis states as
follows: $\widehat{n}\left\vert 0\right\rangle =0$ and $\widehat{n}%
\left\vert 1\right\rangle =\left\vert 1\right\rangle $.

At this point we assume that the qubit's dynamics is much faster than that
of the classical NR, so the equation for the NR can be averaged over the
period $2\pi /\omega $ and then the NR's dynamics is defined by the
time-averaged voltage
\begin{equation}
\overline{V}_{\mathrm{I}}=\frac{2e(n_{\mathrm{g}}-\overline{\left\langle
n\right\rangle })}{C_{\Sigma }}.  \label{VIave}
\end{equation}%
In what follows this time averaging is assumed.

Denoting the sum of the constant terms in Eqs.~(\ref{F_GNR}, \ref{F_NR}) as $%
F_{0}$, we obtain%
\begin{equation}
F=F_{0}+\frac{\partial F}{\partial x}x-F_{\mathrm{A}}\cos \omega _{\mathrm{rf%
}}t,
\end{equation}%
\begin{equation}
\frac{\partial F}{\partial x}=-\frac{2}{C_{\Sigma }}\left( \frac{C_{\mathrm{%
NR}}V_{\mathrm{NR}}}{\xi }\right) ^{2}\left[ 1-\frac{\partial \left\langle
n\right\rangle }{\partial n_{\mathrm{g}}}\right] .  \label{dFdx}
\end{equation}%
The term $F_{0}$ results in an (irrelevant) constant displacement of the NR,
while the linear term results in the resonance frequency shift in Eq.~(\ref%
{eq4x}) as follows%
\begin{equation}
m\omega _{0}^{2}-\frac{\partial F}{\partial x}\equiv m\widetilde{\omega }_{%
\mathrm{NR}}^{2}.
\end{equation}%
Then we obtain the NR's frequency shift
\begin{equation}
\Delta \widetilde{\omega }_{\mathrm{NR}}=\widetilde{\omega }_{\mathrm{NR}%
}-\omega _{0}\approx \frac{1}{2m\omega _{0}}\frac{\partial F}{\partial x}%
\equiv \Delta \omega _{1}+\Delta \omega _{2},
\end{equation}%
where $\Delta \omega _{1}$ and $\Delta \omega _{2}$ correspond to the two
terms in Eq.~(\ref{dFdx}). The term $\Delta \omega _{1}$ does not depend on
the state of the qubit; we therefore define the qubit-state-dependent
frequency shift%
\begin{equation}
\Delta \omega _{\mathrm{NR}}=\Delta \widetilde{\omega }_{\mathrm{NR}}-\Delta
\omega _{1}=\Delta \omega _{2}  \label{DwNR}
\end{equation}%
which leads to Eq.~(\ref{DwNR_1}).

\section{Quantum capacitance}

In addition to the theory presented in the previous Appendix, it
is useful to consider the system qubit-resonator by introducing
the quantum capacitance, which is the subject of this Appendix.
\begin{figure}[t]
\includegraphics[width=7.5 cm]{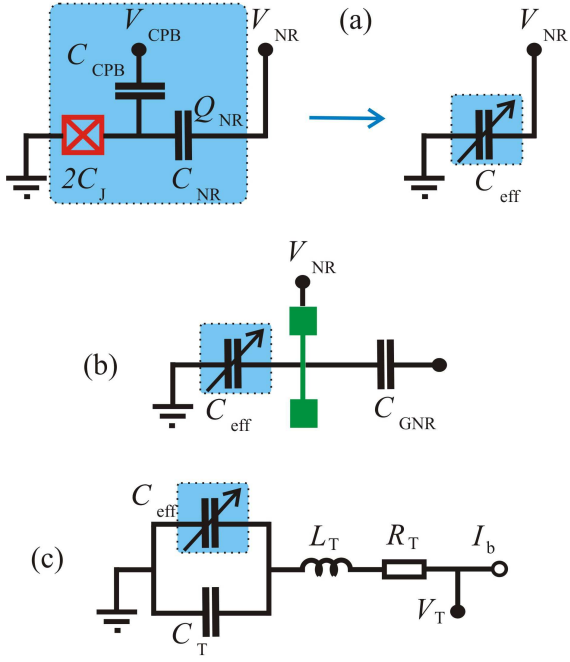}
\caption{(Color online) Scheme showing how the charge qubit can be
described as an effective capacitance coupled either to the NR or
to $LCR$ resonator. (a) To the left, the charge qubit (CPB) is
shown to be described as the capacitance $2C_{\mathrm{J}}$
controlled by the voltage $V_{\mathrm{CPB}}$ and coupled through
the coupling capacitance $C_{\mathrm{NR}}$ to a
measuring circuitry. This is described as the effective capacitance $C_{%
\mathrm{eff}}$ as shown to the right. (b) The effective
capacitance is
coupled to the NR, which can be used to model our system shown in Fig.~%
\protect\ref{Fig:scheme}. (c) The effective capacitance is coupled
to the electric $LCR$ tank circuit.} \label{Fig:Scheme2}
\end{figure}

Let us introduce the effective (differential) capacitance, as it is shown in
Fig.~\ref{Fig:Scheme2}(a), by differentiating the charge $Q_{\mathrm{NR}}$
of the capacitance $C_{\mathrm{NR}}$ as follows\cite{Johansson06}: $C_{%
\mathrm{eff}}=\partial Q_{\mathrm{NR}}/\partial V_{\mathrm{NR}}$. Then, for
the charge $Q_{\mathrm{NR}}=(V_{\mathrm{NR}}-\overline{V}_{\mathrm{I}})C_{%
\mathrm{NR}}$ with the island's voltage given by Eq.~(\ref{VIave}), we obtain%
\begin{equation}
C_{\mathrm{eff}}=C_{\mathrm{geom}}+C_{\mathrm{Q}},  \label{Ceff}
\end{equation}%
which consists of the quantum capacitance $C_{\mathrm{Q}}$, given by Eq.~(%
\ref{CQ1}), and the geometric capacitance $C_{\mathrm{geom}}$\
\begin{equation}
C_{\mathrm{geom}}=\frac{C_{\mathrm{NR}}(C_{\Sigma }-C_{\mathrm{NR}})}{%
C_{\Sigma }}\approx \frac{2C_{\mathrm{J}}C_{\mathrm{NR}}}{2C_{\mathrm{J}}+C_{%
\mathrm{NR}}},\text{ }  \label{Cgeom}
\end{equation}%
where the latter approximation is valid for $C_{\mathrm{CPB}}\ll C_{\mathrm{J%
}},C_{\mathrm{NR}}$.

Alternatively to the approach of the previous Appendix, one can consider the
force $F_{\mathrm{NR}}$ as the electrostatic force from the effective
capacitance [see Fig.~\ref{Fig:Scheme2}(b)]: $F_{\mathrm{NR}}=\frac{1}{2}%
\frac{\partial }{\partial x}\left( C_{\mathrm{eff}}V_{\mathrm{NR}%
}^{2}\right) $. Then the term with the quantum capacitance, in which $C_{%
\mathrm{NR}}^{2}\approx C_{\mathrm{NR}0}^{2}\left( 1+x/\xi \right) ^{2}$,
results in the same frequency shift as obtained in the previous Appendix,
Eq.~(\ref{DwNR}).

\begin{figure}[b]
\includegraphics[width=7.5cm]{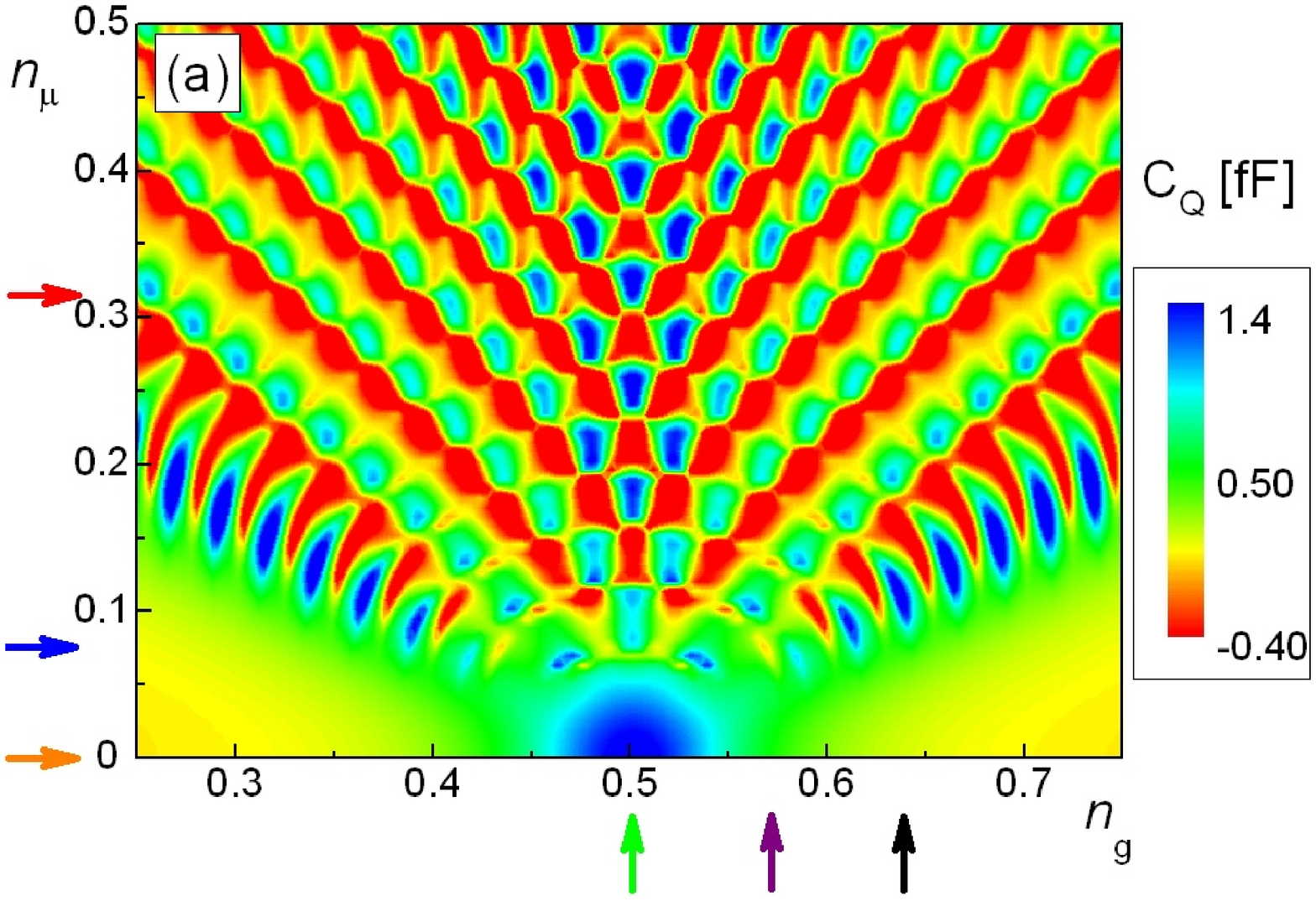} \includegraphics[width=8cm]{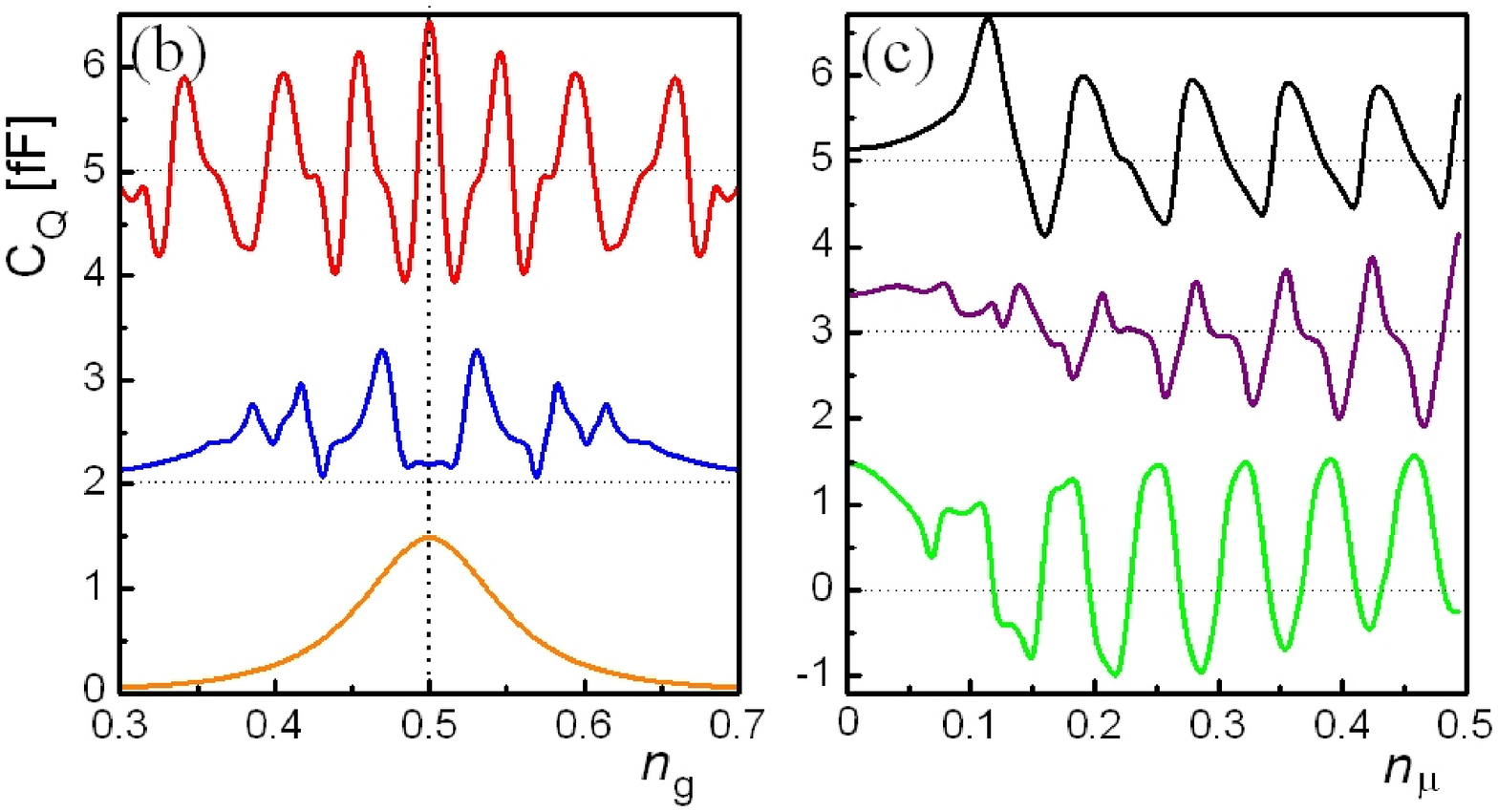}
\caption{(Color online) LZS interferometry probed via a quantum
capacitance.
(a) The quantum capacitance $C_{Q}$ of the qubit versus the energy bias ($n_{%
\mathrm{g}}$) and the driving amplitude ($n_{\protect\mu }$).
Arrows show the values of $n_{\protect\mu }$ and $n_{\mathrm{g}}$
at which the graphs
(b) and (c) are plotted as functions of $n_{\mathrm{g}}$ and $n_{\protect\mu %
}$, respectively. The upper curves were shifted for clarity.}
\label{Fig:CQ}
\end{figure}

\section{Qubit probed by tank circuit}

In this Appendix we consider a qubit coupled capacitively to the series $LCR$
(tank) circuit [see Fig.~\ref{Fig:Scheme2}(c)]. The tank circuit consists of
an inductor $L_{\mathrm{T}}$ and a capacitor $C_{\mathrm{T}}$, while
dissipation is described by the resistor $R_{\mathrm{T}}$. The qubit is
considered to be coupled to the tank circuit through the coupling
capacitance, which for uniformity we again denote by $C_{\mathrm{NR}}$ (even
though there is no NR in the scheme considered in this Appendix), in
parallel to the tank's capacitance $C_{\mathrm{T}}$. The effect of the qubit
on the tank circuit can be described by replacing the tank capacitance $C_{%
\mathrm{T}}$ with $\widetilde{C}_{\mathrm{T}}=C_{\mathrm{T}}+C_{\mathrm{eff}%
} $, where the effective capacitance of the Cooper-pair box is given by Eq.~(%
\ref{Ceff}). The geometric capacitance $C_{\mathrm{geom}}$ gives only a
constant contribution to the tank capacitance $C_{\mathrm{T}}$, while the
quantum capacitance $C_{\mathrm{Q}}\ll C_{0}=C_{\mathrm{T}}+C_{\mathrm{geom}%
} $ is defined by the derivative of the average extra Cooper-pair
number on the island $\left\langle n\right\rangle $.

The tank circuit is biased by the current $I_{\mathrm{b}}=I_{A}\cos \omega _{%
\mathrm{rf}}t$. The output voltage is given by $V_{\mathrm{T}}=V_{A}\cos
(\omega _{\mathrm{rf}}t+\theta )$. Then from the equation for the voltage we
obtain for the phase shift%
\begin{equation}
\tan \theta =Q_{0}\left( 2\frac{\Delta \omega }{\omega _{0}}+\frac{C_{%
\mathrm{Q}}}{C_{0}}\right) ,  \label{tantheta}
\end{equation}%
\begin{equation}
\omega _{0}=\frac{1}{\sqrt{L_{\mathrm{T}}C_{0}}}\text{, }\Delta \omega
=\omega _{\mathrm{rf}}-\omega _{0}\text{, }Q_{0}=\frac{1}{R_{\mathrm{T}}}%
\sqrt{\frac{L_{\mathrm{T}}}{C_{0}}}.
\end{equation}%
The measured value can be either the voltage shift $\theta $ at resonance
frequency ($\Delta \omega =0$)\cite{Duty05, Sillanpaa05, Paila09}%
\begin{equation}
\tan \theta =Q_{0}\frac{C_{\mathrm{Q}}}{C_{0}},
\end{equation}%
or the resonance frequency shift (at which the voltage shift $\theta =0$):%
\cite{LaHaye09}%
\begin{equation}
\frac{\Delta \omega }{\omega _{0}}=-\frac{C_{\mathrm{Q}}}{2C_{0}}.
\end{equation}%
Both are proportional to the quantum capacitance $C_{\mathrm{Q}}$.

For the sake of illustration, in addition to Fig.~\ref{Fig:Dw}, we also
demonstrate in Fig.~\ref{Fig:CQ} the direct LZS interferometry calculated
for the quantum capacitance for the parameters of Ref.~[%
\onlinecite{Sillanpaa06}]: $E_{\mathrm{J}0}/h=12.5$ GHz, $E_{\mathrm{C}%
}/h=24 $ GHz, $\omega /2\pi =4$ GHz, $k_{B}T/h=1$ GHz, and also we
have taken $\alpha =0.005$, $B=0.5$. We note that besides the
difference in the parameters, in Fig.~\ref{Fig:Dw} the frequency
shift $\Delta \omega $ was plotted, while in Fig.~\ref{Fig:CQ} the
quantum capacitance $C_{\mathrm{Q}}$ was shown. Both figures were
calculated by numerically solving the Bloch equation.

Finally, it is worthwhile emphasizing that for simplicity we have assumed
that the qubit's dynamics is much faster than the resonator's dynamics. In
the general case, the cooperative dynamics of the qubit-resonator system
should be studied, as e.g. in Ref.~[\onlinecite{Greenberg08}]. However, a
simplification can be made because the stationary oscillations in the
\textit{nonlinear} system (either NR or tank circuit), influenced by the
qubit's dynamics, can be reduced to oscillations in the \textit{linear}
system, as was studied in Ref.~[\onlinecite{Shevchenko08}]. In that work,
the Krylov-Bogolyubov technique of asymptotic expansion was used. This
technique describes the influence of the qubit as shifts of both the
effective damping factor and the effective coefficient of elasticity. In
analogy to the results of Ref.~[\onlinecite{Shevchenko08}], for the system
considered here, this means that not only the voltage shift $\theta $ is
related to the qubit's capacitance $C_{\mathrm{Q}}$ [see Eq.~(\ref{tantheta}%
)], but also the voltage magnitude $V_{A}$\ is defined by $C_{\mathrm{Q}}$.
This, in particular, explains the experimental results presented in Fig.~3
by Paila~\textit{et~al.}~[\onlinecite{Paila09}].

\end{document}